\documentclass[manuscript]{aastex}
\usepackage{amsmath}
\usepackage{txfonts}
\usepackage{booktabs}
\usepackage[pagewise]{lineno}

\slugcomment{Not to appear in Nonlearned J., 45.}
\shorttitle{MHD waves in active region corona}
\shortauthors{Kitagawa, Yokoyama, Imada, \& Hara}
\begin{document}
    \title{Mode identification of MHD waves in an active region observed with Hinode/EIS}
    \author{N. Kitagawa and T. Yokoyama}
    \affil{Department of Earth and Planetary Science, University of Tokyo, 7-3-1 Hongoh, Bunkyoh-ku, Tokyo 113-0033, Japan}
    \email{kitagawa@eps.s.u-tokyo.ac.jp}
    \author{S. Imada}
    \affil{Institute of Space and Astronautical Science, Japan Aerospace Exploration Agency, 3-1-1 Yoshinodai, Chuo-ku, Sagamihara-shi, Kanagawa 252-5210, Japan}
    \and
    \author{H. Hara}
    \affil{National Astronomical Observatory of Japan, 2-21-1 Osawa, Mitaka-shi, Tokyo 181-8588, Japan}
\begin{abstract}
	In order to better understand the possibility of coronal heating by MHD waves, 
	we analyze Fe \textsc{xii}  $195.12\mspace{3mu}${\AA} data observed with EUV Imaging Spectrometer (EIS) onboard \textit{Hinode}. 
	We performed a Fourier analysis of EUV intensity and Doppler velocity time series data in the active region corona. 
    Notable intensity and Doppler velocity oscillations were found for two moss regions out of the five studied, 
	while only small oscillations were found for five apexes of loops.
	The amplitudes of the oscillations were $0.4-5.7\mspace{3mu}${\%} for intensity 
	and $0.2-1.2\mspace{3mu}\mathrm{km}\mspace{3mu}\mathrm{s}^{-1}$ for Doppler velocity.
	In addition, oscillations of only Doppler velocity 
	were seen relatively less often in the data. 
     We compared the amplitudes of intensity and those of Doppler velocity in order to identify MHD wave modes, 
	and calculated the phase delays between Fourier components of intensity and those of Doppler velocity.
     The results are interpreted in terms of MHD waves as follows:  
    (1) few kink modes or torsional Alfv\'{e}n mode waves were seen in both moss regions and the apexes of loops; 
    (2) upwardly propagating and standing slow mode waves were found in moss regions; 
    and (3) consistent with previous studies, 
	estimated values of energy flux of the waves were several orders of magnitude lower than that required for heating active regions. 
\end{abstract}
\section{Introduction}
    There is much disagreement over which mechanisms heat the solar atmosphere above the chromosphere up to the coronal temperature of several million Kelvin. 
    At present, two major possibilities are often suggested: Alfv\'{e}n wave heating and nanoflare heating. 
    However, waves with sufficient energy flux to heat the corona have not yet been detected in the corona, except for waves triggered by flares. 
    Furthermore, the typical spatial scale of nanoflares is often considered to be smaller than the resolution of the current telescopes. 
    It is also uncertain which mechanism is dominant, or whether only one mechanism is dominant. 
    Although a large number of wave and oscillation events have been observed \citep{asc04}, 
	to further understand coronal heating, 
	it is necessary to analyze the statistical properties of waves in the corona, 
    such as the relation between oscillation phenomena and their locations, 
	and between MHD modes and their locations. 
	\citet{asc02} conducted a statistical analysis of the physical parameters of transverse loop oscillations 
	observed with the \textit{Transition Region And Coronal Explorer} (TRACE). 
    They investigated 26 coronal loops with periods of $2-11\mspace{3mu}\mathrm{min}$ (decay time: $3-21 \mspace{3mu} \mathrm{min}$) 
	and concluded that the oscillations could be described not by the simple kink eigen mode 
    oscillations but by flare-induced impulsively generated MHD waves. 
    In another study, Doppler shift oscillations in hot coronal loops were investigated using the Solar Ultraviolet Measurement of Emitted Radiation (SUMER), 
    in which 54 oscillations in 27 flare-like events were analyzed \citep{wan03}. 
    The periods of oscillations were $7-31 \mspace{3mu} \mathrm{min}$ (decay time: $6-37 \mspace{3mu} \mathrm{min}$), 
    which were somewhat longer than the TRACE loop oscillations. 
    The authors interpreted these waves in terms of slow magnetoacoustic standing waves in hot loops 
	because the phase speeds roughly agreed with the sound speed and the phase delays of intensity between Doppler shift were 1/4 period. 
	Observations of coronal oscillatory phenomena were also made with the EUV Imaging Spectrometer (EIS) onboard \textit{Hinode}, 
	with the data suggesting the existence of 
	slow mode \citep{mar08, wan09a, wan09b}, slow mode and fast sausage mode \citep{erd08}, and fast kink mode \citep{van08}. 
    In addition, recent studies statistically showed the existence of propagating slow magnetoacoustic waves in coronal holes using SUMER  and EIS data \citep{ban09,gup09}. 
    Finally, \citet{osh09} conducted a statistical analysis of intensity oscillations in an active region using consecutive $40''$ slot images of 
	an Fe \textsc{xii} $195.12\mspace{3mu}${\AA} emission line observed with \textit{Hinode}/EIS. 
    They found that oscillations with frequencies in the range of $2-8\mspace{3mu}\mathrm{mHz}$ were concentrated in bright plages, 
    while oscillations of higher frequencies were concentrated in the edges of those areas. 
    The oscillations were interpreted in terms of magnetoacoustic waves at frequencies less than 8 mHz and fast magnetoacoustic waves at frequencies above that. 
    However, their data did not have information about line profiles, so the identification of MHD waves was not conclusive. 

	In order to reveal the nature of MHD waves in the corona, we need to take into account the properties of coronal structures such as loops or moss. 
	Recent observations of active region loops with EIS have revealed that coronal loops are not isothermal, 
	and that their filling factors are approximately $10$\%, 
	which supports the existence of multi-strands of sub-resolution size \citep{war08}.
    Moss has a bright reticulated pattern, which emits EUV by thermal conduction from overlying hot and relatively high-pressure loops \citep{ber99}. 
    The time scale of moss dynamics is $10-30\mspace{3mu} \mathrm{s}$, 
    which is due to internal intensity variation, 
	obscuration by chromospheric jets, 
	or mechanical interaction with those jets \citep{dep99}. 
    \citet{kat05} found that filling factors are lower at the footpoints of hot loops than at the footpoints of cool loops. 
    They concluded that braiding of the coronal magnetic field is more efficient at hot loops than at cool loops, 
	which means there is a larger magnetic energy release at loops with a lower filling factor. 
    Recently, an Fe \textsc{xii} $195\mspace{3mu}${\AA} emission line observed with \textit{Hinode}/EIS was analyzed 
	to investigate the time variability of intensity, Doppler velocity, and nonthermal velocity in moss \citep{bro09}. 
    There was less than 15\% variability in intensity, $3 \mspace{3mu} \mathrm{km} \mspace{3mu} {\mathrm{s}}^{-1}$ variability in Doppler velocity, 
	and the average nonthermal velocity was $21.7 \mspace{3mu} \mathrm{km} \mspace{3mu} {\mathrm{s}}^{-1}$. 

    In this study, we analyzed time series data of intensity and Doppler velocity for loops and moss in an active region 
	using sit-and-stare mode data observed with EIS \citep{cul07}. 
    Our purpose is to identify the modes of detected oscillations in this active region 
	in order to understand which modes of MHD waves contribute to the heating of active regions. 
    We compared Fourier amplitudes of intensity and Doppler velocity to determine the modes of the detected oscillatory signals, 
    and then investigated the correspondence between these coronal structures and the modes. 
\section{Data reduction and analysis}
    We used sit-and-stare mode data observed with EIS. 
    The duration of the data was 2006 Dec 3 12:43:10-13:37:20 UT, 
    and the field of view was $1''$ in the $x$-direction and $256''$ in the $y$-direction in heliocentric coordinates. 
    There were 288 data points in the time direction. 
    The exposure time was $10 \mspace{3mu} \mathrm{s}$, 
	and the average time cadence was $11.3\mspace{3mu} \mathrm{s}$, with a range of $11.3-11.5\mspace{3mu}\mathrm{s}$.
    The standard Solar Software interactive data language (SSWIDL) procedure \texttt{eis\_prep} was used to calibrate the data. 
    We used an Fe \textsc{xii} $195.12\mspace{3mu}${\AA} emission line, whose line profile at each pixel was fitted using a single Gaussian function. 
    Then intensity (integrated line profile) and Doppler shift (difference between line centroid and 
    $195.12\mspace{3mu}${\AA)} were calculated using the fitted parameters. 
	After removing the influence of the slit-tilt and the orbital variation using the \texttt{eis\_wave\_corr} procedure, 
	we were able to investigate the average line centroid at each exposure. 
	However, a $90-100\mspace{3mu} \mathrm{min}$ period component still remained, 
	which was subtracted for accuracy.
	Usually, the zero-point of Doppler velocity is removed through examination of a quiet sun region, 
	where Doppler velocity is much smaller and less dynamic. 
	However, with a short exposure time ($10\mspace{3mu}\mathrm{s}$) in our data, 
	the incident photons at the quiet sun region were too small for analysis, 
	so the zero-point of Doppler velocity was calculated by using all pixels along the slit. 
	Since no explosive events were observed in our data, this zero-point adjustment does not produce artificial oscillations.
	Note that there were no prominent wing components in the line profiles, 
    and the weak line blend with Fe \textsc{xii} $195.18\mspace{3mu}${\AA} \citep{you09} can be ignored 
    because the main focus of our research is the time variability of Doppler velocity \citep{wan09b}.

\begin{figure}
    \includegraphics[width=16cm,clip]{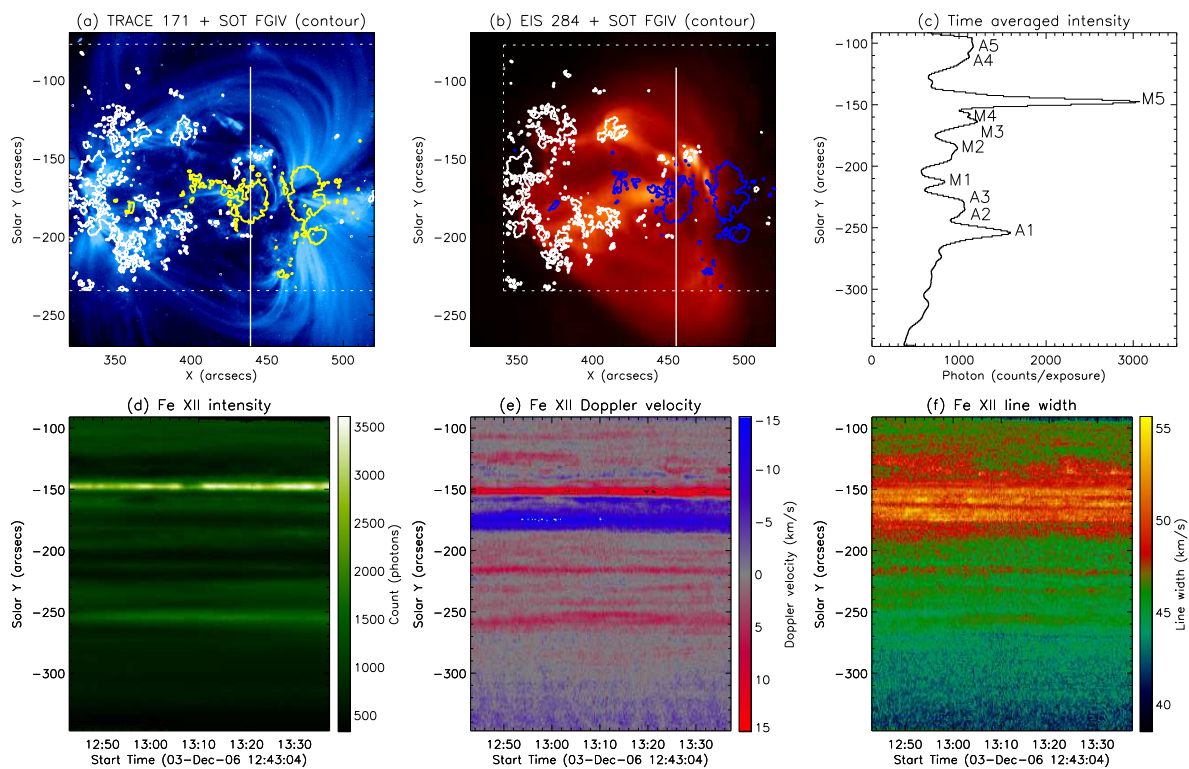}
    \caption{(a) TRACE $171\mspace{3mu}${\AA} image obtained at 13:36:46UT on 2006 Dec 3. 
                 (b) EIS Fe \textsc{xv} $284.16\mspace{3mu}${\AA} raster scan image obtained at 15:32:19-17:46:31UT on 2006 Dec 3. 
                 In panels (a) and (b), the white vertical line indicates the cotemporal position of the EIS slit, 
				 and Fe \textsc{i} $6302\mspace{3mu}${\AA} Stokes-V data from \textit{Hinode}/SOT are also shown. 
                 White and yellow/blue contours in the two images indicate positive and negative magnetic fields, respectively, 
                 and SOT's field of view is shown with white dashed lines.
                 (c) Time averaged intensity at each pixel on the EIS slit. 
                 The lower panels show time series data of the EIS sit-and-stare mode observation starting at 12:43:10UT on 2006 Dec 3. 
                 The time cadence for each exposure is $11.4 \mspace{3mu} \mathrm{s}$. 
                 (d) Intensity. 
                 (e) Doppler velocity. 
                 (f) Line width. 
                 }
    \label{ss trace}
\end{figure}

    Panel (a) in Figure \ref{ss trace} shows a $171\mspace{3mu}${\AA} image of NOAA active region 10926 taken by TRACE. 
    Panel (b) in Figure \ref{ss trace} shows an EIS Fe \textsc{xv} $284.16\mspace{3mu}${\AA} raster scan image.  
    The white vertical line in each panel indicates the cotemporal position of the EIS slit in both panels. 
    In the TRACE image, we can see that the slit cuts across some apexes of coronal loops and some moss regions. 
	Fe \textsc{i} $6302\mspace{3mu}${\AA} Stokes-V data from \textit{Hinode}/SOT are also shown in panels(a) and (b). 
	White and yellow/blue contours in the two images indicate positive and negative magnetic fields, respectively. 
    SOT's field of view is represented by a white dotted rectangular area. 
    Panel (c) in Figure \ref{ss trace} shows the time averaged intensity at each pixel. 
    We analyzed the structures which displayed a high intensity value (i.e., a high signal-to-noise ratio),
    and considering the correspondence between the structures seen in images (a) and (b) and the intensity profile of the EIS slit in panel (c), 
    we identified the structures of the studied locations. 
    A1-A5 and M1-M5 in panel (c) represent apexes of loops and moss regions, respectively. 
    A small number of pixels in the top and bottom of the slit were excluded from the analysis due to the effect of a jitter-removing process, 
	as outlined below. 

    To remove the influence of a jitter of the spacecraft, we used data from the Ultra Fine Sun Sensor (UFSS). 
    After subtracting the jitter in the N-S direction from the original time series data, 
	a trend with a time scale of $\sim$10 min in the $y$-direction still remained, 
	similar to that observed by \citet{utz10}. 
    Then we calculated the cross-correlation coefficient between one shot and its following shot in the original data 
	and obtained the shift in the $y$-direction where the coefficient was maximized. 
	The absolute values of the shift from the cross-correlation method were somewhat larger than those of the UFSS output in the N-S direction. 
    We shifted the original data using both shift values, and concluded that the cross-correlation method was more suitable. 
    Using this shift process, we could remove apparent oscillations in the $y$-direction caused by the jitter. 
	The UFSS output in the E-W direction ($x$-direction) showed a maximum drift of $2''$. 
	Because this is comparable to the spatial resolution of the EIS $1''$ slit, which approximately equals $2''$ \citep{har08}, 
	we neglected the influence of the jitter in the $x$-direction. 
    The processed data are shown in panels (d), (e), and (f) in Figure \ref{ss trace}. 

\begin{figure}
    \centering
    \includegraphics[width=16cm,clip]{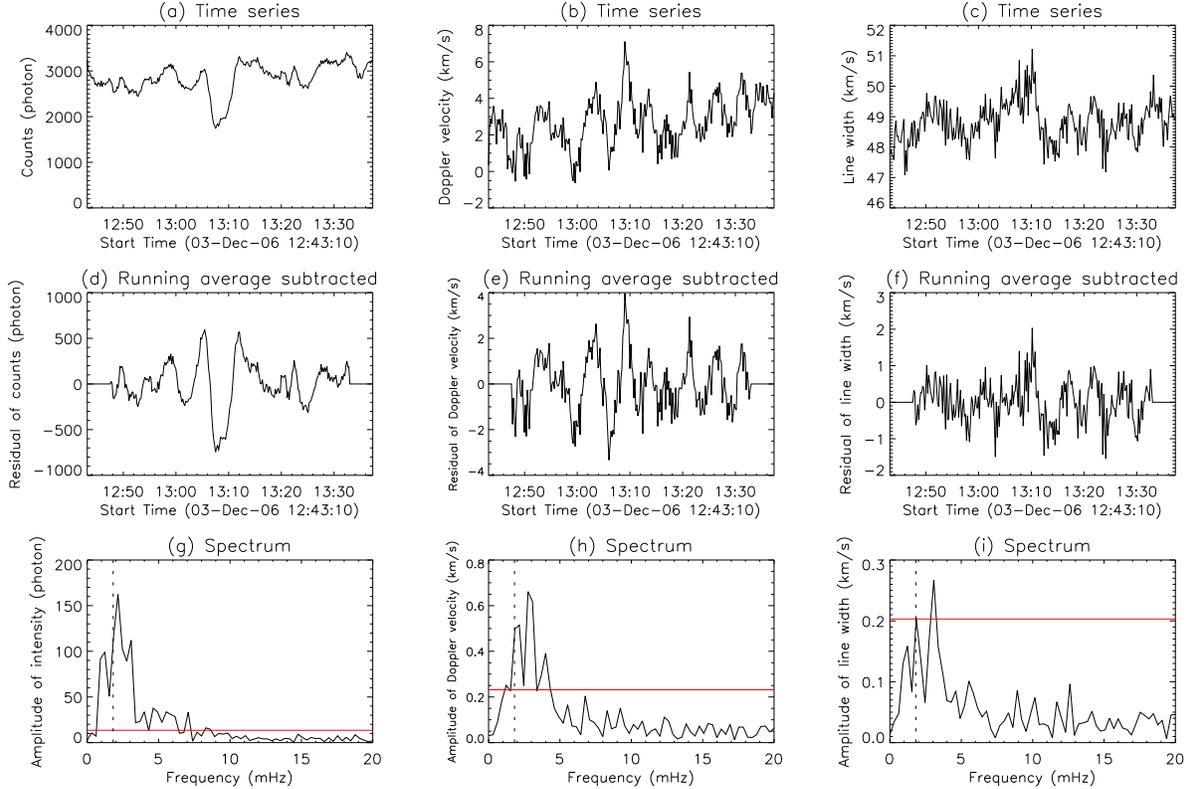}
    \caption{Examples of time series and Fourier spectra 
                 (at $y=-147''$ in the EIS slit; M5). 
                 (a), (b), and (c): Time series of intensity, Doppler velocity, and line width, respectively. 
                 (d), (e), and (f): Residuals after subtracting a 48-point running average from the original time series data. 
                 (g), (h), and (i): Fourier spectra.
					The vertical dotted line in each plot indicates an artificial low-frequency cut-off of 1.8 mHz (540 s).
					The horizontal red line in each plot indicates the 99\% significance level calculated from Poisson noise of 
					incident photons.}
    \label{time series}
\end{figure}

    Time series data and Fourier spectra for intensity, Doppler velocity, and line width 
	at $y=-147''$ (M5) in the slit are shown in Figure \ref{time series}. 
	The time series data for each pixel was interpolated on to a constant $11.3\mspace{3mu}\mathrm{s}$ cadence grid, 
	before being subjected to rigorous Fourier transforms. 
    Panels (a), (b), and (c) show the time series of intensity, Doppler velocity, and line width, respectively. 
    Before calculating the Fourier transform, we subtracted a 48-point running average from the original time series data
    in order to reduce the influence of trends longer than, or as long as, the duration of the data, 
    which resulted in a decrease in power for periods longer than $48 \times 11.3 \simeq 540 \mspace{3mu}\mathrm{s}$. 
    This introduced an artificial low-frequency cut-off of $1.8 \mspace{3mu} \mathrm{mHz}$, resulting in detections below this value being less reliable. 
    The time series data after the subtraction process are shown in panels (d), (e), and (f). 
    We defined oscillations as signals which exceed the 99\% significance level calculated by assuming the Poisson noise of incident photons. 
    This is indicated by a red line in panels (g), (h), and (i). 
    
\section{Results}
    In Figure \ref{wmst}, pairs of Fourier components with the same frequency $(\tilde{I}(\nu)/I_0,\tilde{v}\mspace{3mu}(\nu)/c_S)$ are plotted, 
    with the horizontal axis in each plot indicating intensity amplitude ($\tilde{I}\mspace{3mu}$) normalized by the time averaged intensity ($I_0$) at each pixel, 
    and the vertical axis in each plot indicating Doppler velocity amplitude ($\tilde{v}$) normalized by the sound speed 
    ($c_S \mspace{-3mu} \sim \mspace{-3mu} 167 \mspace{3mu} \mathrm{km} \mspace{3mu} {\mathrm{s}}^{-1}$) 
    at the formation temperature of Fe \textsc{xii} ($\sim \mspace{-3mu} 1.3 \mspace{3mu} \mathrm{MK}$). 
    The symbols in each plot represent data points where: 
			only intensity amplitude is significant (red dots), 
			only Doppler velocity amplitude is significant (blue crosses), 
			both intensity and Doppler velocity amplitude are significant (green diamonds), 
			and neither intensity nor Doppler velocity amplitude are significant (black dots). 
	Red triangles represent data points where both intensity and line width amplitudes exceed the significance level, 
	while red dots indicate data points where line width amplitude is not significant. 
	The three orange dotted lines show the relationship between intensity amplitude and Doppler velocity amplitude for magneto-acoustic waves 
	when the angle between the magnetic field and the line of sight is 0, 60, and 75 degrees. 
	From Figure \ref{wmst}, we found that there were only a few significant oscillations at the apexes of loops, 
	while there were a substantial amount of significant intensity and Doppler velocity oscillations at the two moss regions. 

    According to MHD wave theory, there are four major modes of magnetic tubes in the corona: 
	torsional Alfv\'{e}n mode, fast kink mode, fast sausage mode, and slow mode \citep{rob84, spr82}. 
    Torsional Alfv\'{e}n mode is incompressible in the first order, so it produces no density (i.e., intensity) oscillations. 
    Assuming our observation resolves both the spatial scale of the torsional Alfv\'{e}n mode and the temporal period of the mode, 
	then out-of-phase Doppler velocity oscillations can be observed at opposite edges of the waveguide.
    However, if either the spatial scale or the temporal period of the mode are unresolved, 
	emission line broadening will be observed instead of Doppler velocity oscillations. 
	Transversal oscillations of loops occur in fast kink mode, which means only Doppler velocity oscillations are likely to be observed as a marker. 
    In addition, intensity oscillations are not observed because of the incompressibility of fast kink mode in the first order. 
    In fast sausage mode, radial oscillations of loops produce intensity oscillations. 
    Since the two surfaces of a loop along the line of sight oscillate in opposite directions and cancel each other out, 
    emission line broadening is observed instead of Doppler velocity oscillations. 
    Slow mode produces both intensity and Doppler velocity oscillations. 
	Since the solar corona is an optically thin and collisionally excited atmosphere, intensity is proportional to the square of density ($I \propto \rho^2$). 
	By performing a linear approximation, we obtained the relationship between density perturbation $\tilde{\rho}_1$ and intensity perturbation $\tilde{I}_1$,
	\begin{equation}
        \dfrac{\tilde{I}_1}{I_0} = 2 \dfrac{\tilde{\rho}_1}{\rho_0}, 
        \label{I rho}
    \end{equation}
    where $I_0$ is intensity and $\rho_0$ is density in the unperturbed state. 
    Then combining Eq.(\ref{I rho}) with the Fourier transform of the linearized continuity equation, we obtained 
    \begin{equation}
        \dfrac{{\tilde{I}}_1}{2 I_0} = \dfrac{{\tilde{v}}_1}{ c_{ \mathrm{ph} } }, 
    \end{equation}
    where $\tilde{v}_1$ is velocity perturbation, and phase speed $c_{\mathrm{ph}}$ is defined as $c_{\mathrm{ph}} \equiv \omega/k$, 
	where $\omega$ is frequency and $k$ is wavenumber. 
	For the slow mode ($c_{ \mathrm{ph} } = c_S$), this relationship is indicated by the orange dotted line ($\theta = 0$) in each plot of Figure \ref{wmst}. 

    Considering the properties of the four MHD modes discussed above, we can interpret Figure \ref{wmst} 
	in a straightforward manner (see Table \ref{modes}). 
    Red triangles (significant oscillations in intensity and line width) can be interpreted as fast sausage mode, 
    and blue crosses (significant oscillations only in Doppler velocity) can be interpreted as torsional Alfv\'{e}n mode or fast kink mode. 
    Green diamonds (significant oscillations in intensity and Doppler velocity) correspond to slow mode, 
	and red dots (significant oscillations only in intensity) may be interpreted as 
	fast sausage mode when line width oscillations are obscured by turbulent plasma motion, 
	or slow mode when Doppler velocity signals are reduced due to the inclination of the magnetic field to the line of sight.
    Finally, black dots represent non-significant oscillations, 
	oscillations of a shorter period than the time cadence, and oscillations of sub-resolution spatial scale.
    The above interpretations therefore suggest that fast sausage mode and slow mode were observed at the moss regions, 
    while fewer oscillations were observed at the apexes of loops. 
	Distinguishing between fast kink mode and torsional Alfv\'{e}n mode is possible by analyzing the phase difference at opposite edges of 
	coronal structures, but this requires spatial resolution high enough to resolve the elemental magnetic strands.

\begin{figure}
    \includegraphics[width=16cm,clip]{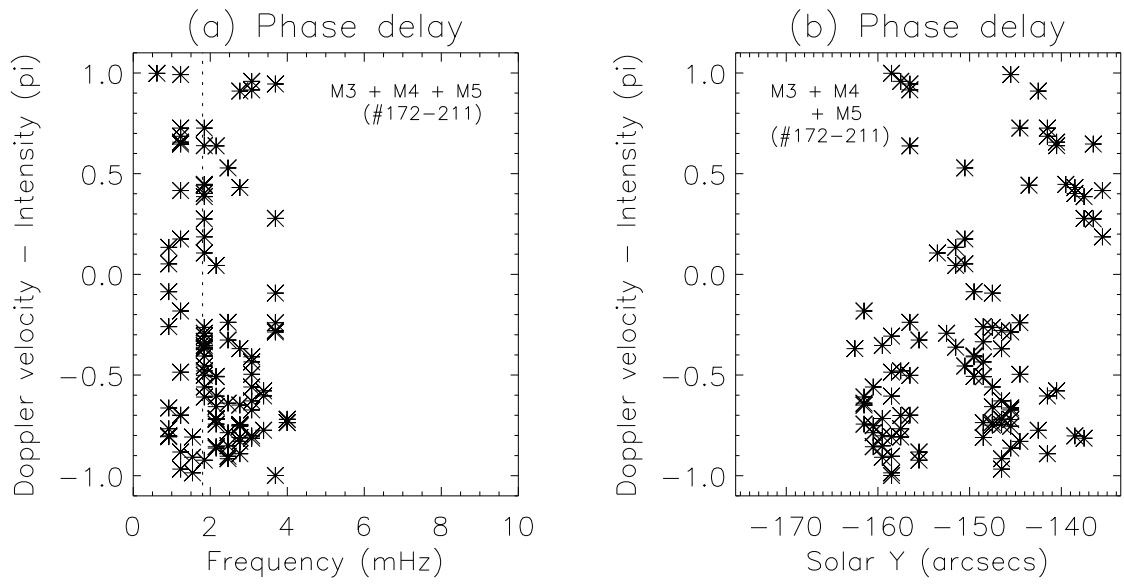}
    \caption{Phase delay between intensity and Doppler velocity at moss regions M3, M4, and M5. 
                 (a) Phase delay as a function of frequency. 
				 The vertical dotted line indicates an artificial low-frequency cut-off of 1.8 mHz (540 s).
                 (b) Phase delay as a function of solar $y$.}
    \label{phst}
\end{figure}

    Phase delays between intensity and Doppler velocity were calculated for data points 
    where both intensity and Doppler velocity amplitudes exceeded the significance level. 
    The results are shown in Figure \ref{phst}. 
    Panel (a) shows frequency in the horizontal axis and phase delay in the vertical axis. 
    Panel (b) shows the position from the bottom of the slit in the horizontal axis and phase delay in the vertical axis. 
    Both panels suggest the existence of upwardly propagating slow mode waves (phase delay: $\pm \pi$) and standing waves (phase delay: $\pm \pi/2$)
    with frequencies of $1-4 \mspace{3mu} \mathrm{mHz}$. 
	Adiabatic theory suggests that phase delays of $0$ or $\pm \pi$ between intensity and Doppler velocity oscillations are indicative of propagating waves, 
	while phase delays of $\pm \pi/2$ suggest the presence of standing waves \citep{deu90}. 
	Phase delays between intensity and Doppler velocity (i.e., V-I phase) are established for all slow mode waves.
\section{Discussion}
    Assuming the coronal loops are resolved by our observations, the absence of waves at the apexes of cool loops means that 
    waves excited by the photospheric granulation or nanoflares must have dissipated before they reached the apexes of cool loops. 
    For the cases in which the spatial scales of the waves were sub-resolution size, incoherency of oscillations may have obscured our observations. 
    This is plausible when considering recent observations which support a multi-strand model \citep{war08}.
    In such cases, non-thermal broadening is enhanced in the transverse direction for torsional Alfv\'{e}n mode, 
    fast kink mode, fast sausage mode, and in the longitudinal direction for slow mode \citep{har99}. 
    In contrast to the observed five apexes, our findings indicate that two moss regions out of the five studied 
	contained large amounts of fast sausage mode and slow mode. 
	Our study revealed intensity oscillations of $0.4-5.7$\% and Doppler velocity oscillations of $0.2-1.2 \mspace{3mu} \mathrm{km} \mspace{3mu} {\mathrm{s}}^{-1}$, 
    with average amplitudes of $1.4$\% and $0.5\mspace{3mu}\mathrm{km}\mspace{3mu}{\mathrm{s}}^{-1}$, respectively. 
    This is consistent with the results of \citet{bro09}. 
    Also detected were upwardly propagating slow mode waves (phase delay: $\pi$ or $-\pi$) with frequencies of $1-4\mspace{3mu} \mathrm{mHz}$. 
    One possible mechanism that excites slow mode is the leakage of 5-min oscillations ($\sim \mspace{-3mu} 3 \mspace{3mu} \mathrm{mHz}$) of the photosphere, 
	as was reported in a study by \citet{dem00}, in which they detected $2-4$\% oscillations in EUV intensity.
    Another possible mechanism is the mechanical interaction between spicules and the surrounding corona \citep{dep03}. 
    If this is the case, frequencies within the $3-5 \mspace{3mu} \mathrm{mHz}$ range might reveal certain properties of spicules, such as lifetime. 
    
\begin{figure}
	\centering
    \includegraphics[width=8cm,clip]{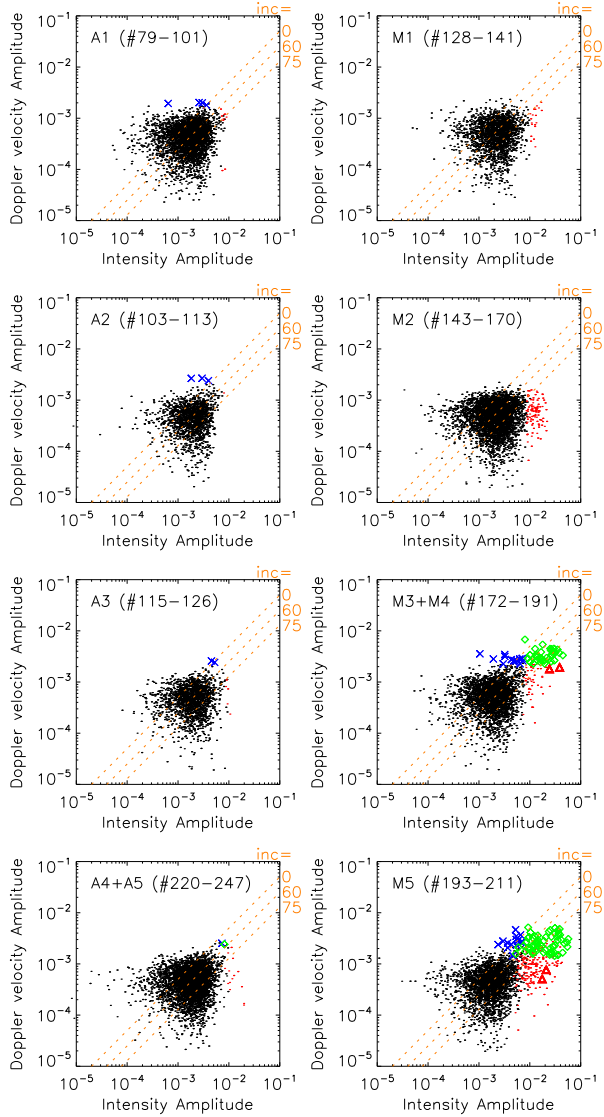}
    \caption{Panels in the left column (A1-A5) show scatter plots at the apexes of TRACE loops, 
                 and panels in the right column (M1-M5) show scatter plots at the moss regions. 
				 The horizontal axis in each plot indicates intensity amplitude normalized by the time averaged intensity. 
                 The vertical axis in each plot indicates Doppler velocity amplitude normalized by the sound speed 
                 ($c_S \mspace{-3mu} \sim \mspace{-3mu} 167 \mspace{3mu} \mathrm{km} \mspace{3mu} {\mathrm{s}}^{-1}$) 
				 at the formation temperature of Fe \textsc{xii} ($\sim \mspace{-3mu} 1.3 \mathrm{MK}$). 
                 Red dots, red triangles, blue crosses, green diamonds, and black dots respectively represent 
                 data points where only intensity amplitude was significant, 
                 both intensity and line width amplitude were significant, 
                 only Doppler velocity amplitude was significant, 
                 both intensity and Doppler velocity Fourier amplitude were significant, 
                 and neither intensity nor Doppler velocity amplitude were significant. 
                 Three orange dotted lines show the relationship between intensity amplitude and Doppler velocity amplitude 
				 for magneto-acoustic waves when the angle between the magnetic field and the line-of-sight is 0, 60, and 75 degrees. 
}
    \label{wmst}
\end{figure}

\begin{figure}
    \includegraphics[width=16cm,clip]{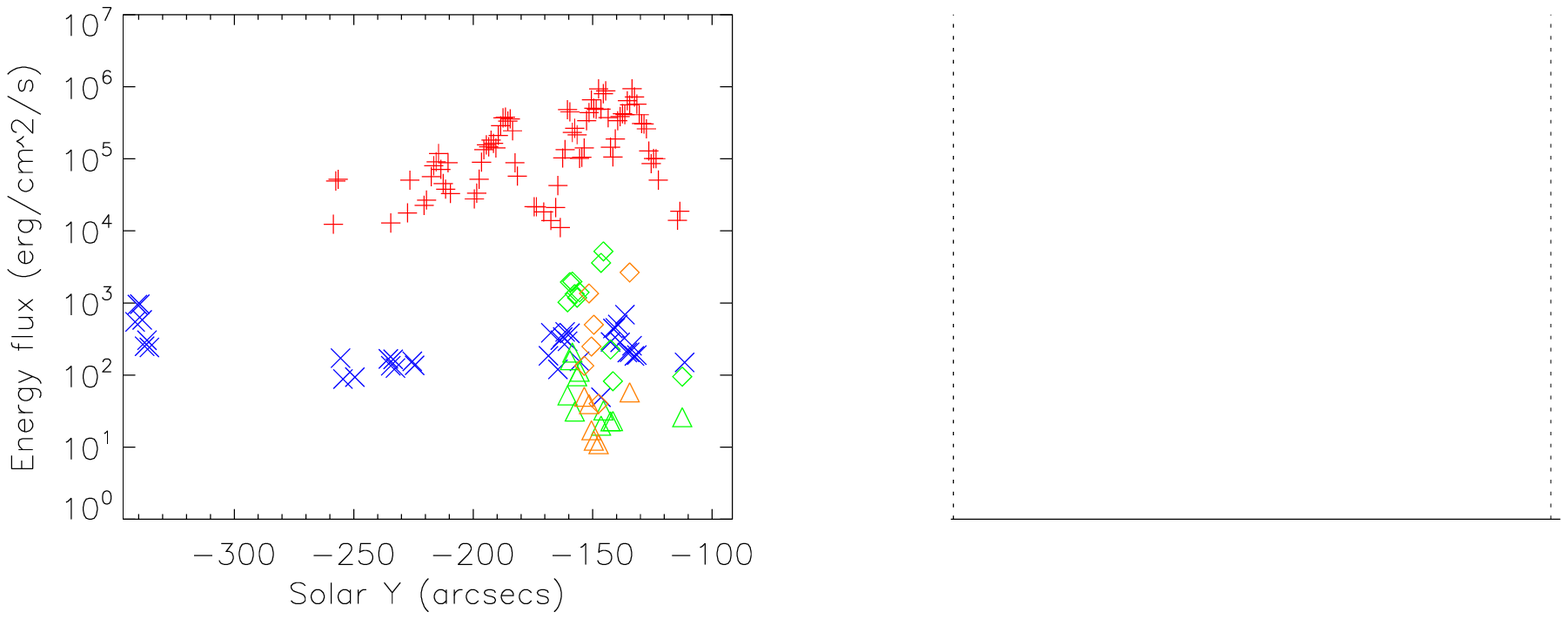}
    \caption{Estimated energy flux in terms of MHD modes. 
				 Red pluses, blue crosses, green diamonds (green triangles), and orange diamonds (orange triangles) respectively represent 
				 energy flux interpreted in terms of fast sausage mode, fast kink mode, 
				 upwardly propagating slow mode calculated by using amplitude of intensity (Doppler velocity), 
				 and downwardly propagating slow mode calculated by using amplitude of intensity (Doppler velocity).
                 }
    \label{power}
\end{figure}

    We crudely estimated the energy flux at each pixel based on the interpretations in terms of MHD waves outlined in Section 3. 
    After combining a linearized continuity equation and Eq. (\ref{I rho}), we obtained 
	\begin{equation}
		\tilde{v}_1 = \dfrac{\omega}{k} \dfrac{\tilde{I}_1}{2 I_0}.  
		\label{v-I}
	\end{equation}
    The line-of-sight effect reduces the observed Doppler velocity $\tilde{v}_{\mathrm{obs}}$ as $\tilde{v}_{\mathrm{obs}} = \tilde{v}_1 \cos \theta$, 
    where $\theta$ is the angle between the direction of the plasma motion and the line of sight. 
    Using Eq. (\ref{v-I}), the energy flux of compressible mode is estimated as follows: 
    \begin{align}
        &\rho_0 [\tilde{I}_1 / (2 I_0)]^2 {c_S}^3  \mspace{18mu} \text{or} \mspace{18mu} 
        \rho_0 \left( \dfrac{\tilde{v}_{\mathrm{obs}}}{\cos \theta} \right)^2 c_S \mspace{36mu} \text{(slow mode), } \\
        &\rho_0 [\tilde{I}_1 / (2 I_0)]^2 {c_A}^3 \mspace{36mu} \text{(fast sausage mode).} 
    \end{align}
	The observed Doppler velocity for fast sausage mode is cancelled along the line of sight at 
	the boundaries of each loop, so we used $\tilde{I}$ instead of $\tilde{v}$ from Eq. (\ref{v-I}). 
    The energy flux transported by torsional Alfv\'{e}n mode or fast kink mode is represented as  
    \begin{equation}
        \rho_0 (\tilde{v}_1)^2 c_A \mspace{36mu} \text{(torsional Alfv\'{e}n or fast kink mode).} \\
    \end{equation}
    The calculated energy flux as a function of position along the slit is shown in Figure \ref{power}. 
    Red pluses and blue crosses represent fast sausage mode and fast kink mode, respectively. 
    Green and orange diamonds respectively represent upwardly and downwardly propagating slow mode estimated using intensity amplitude. 
	As outlined in section 3, we determined propagation direction based on phase delays between intensity and Doppler velocity as follows: 
    upwardly propagating was defined as a phase delay in the range of  $(5/6) \pi$ to $\pi$ and $-\pi$ to $-(5/6)\pi$, 
    while downwardly propagating was defined as a phase delay in the range of $-(1/6)\pi$ to $(1/6)\pi$. 
    Green and orange triangles respectively represent upwardly and downwardly propagating slow mode estimated using Doppler velocity amplitude. 
    We assumed $\rho_0 = 8.4 \times 10^{-16} \mspace{3mu} \mathrm{g} \mspace{3mu} {\mathrm{cm}}^{-3}$, 
    $c_S = 167 \mspace{3mu} \mathrm{km} \mspace{3mu} {\mathrm{s}}^{-1}$, and
    $c_A = 1.0 \times 10^3 \mspace{3mu} \mathrm{km} \mspace{3mu} {\mathrm{s}}^{-1}$. 
    If our interpretations in terms of MHD waves are correct, our findings indicate that 
    fast sausage mode waves have a significant proportion of the energy flux needed for coronal heating in active regions
	($10^7 \mspace{3mu} \mathrm{erg} \mspace{3mu} {\mathrm{cm}}^{-2} \mspace{3mu} {\mathrm{s}}^{-1}$), 
	as estimated by \citet{wit77}. 
    For energy calculations, $\theta$ is assumed to be zero. 
    As a result, derived energies act as lower limits and rise significantly as the inclination angle of the magnetic field increases.
    Note that when Doppler velocity signals are obscured by the inclination of the magnetic field, 
	red dots in Figure \ref{wmst} should be interpreted as a signature of slow mode. 
	This means that the estimated energy flux, shown by red pluses in Figure \ref{power}, is reduced by $2-3$ orders of magnitude 
	as a result of wave energy flux being proportional to the sound speed rather than the Alfv\'{e}n speed.
	Given that all red dots are below the orange line for $\theta = 0$ in Figure \ref{wmst}, 
	a large portion of the red dots may be interpreted in terms of slow mode.
    
    Recent studies have detected torsional Alfv\'{e}n mode in the photosphere \citep{jes09} 
	by investigating line width oscillations of $\sim \mspace{-3mu} 2.6\mspace{3mu}\mathrm{km}\mspace{3mu}\mathrm{s}^{-1}$, 
    fast kink mode in the chromosphere \citep{he09} with velocity amplitudes of $4.7 - 20.8\mspace{3mu}\mathrm{km}\mspace{3mu}\mathrm{s}^{-1}$, 
	and fast kink mode in the corona \citep{tom07} with velocity amplitudes $\sim \mspace{-3mu} 0.3\mspace{3mu}\mathrm{km}\mspace{3mu}\mathrm{s}^{-1}$. 
    Moreover, considering the extremely slow rate of dissipation of torsional Alfv\'{e}n mode and fast kink mode waves, we expected to detect them in our observations. 
    However, there were few signatures of either. 
    There are three possible explanations for this. 
    Firstly, torsional Alfv\'{e}n mode and fast kink mode waves may have been reflected below the corona 
	due to sudden changes 
	in plasma density at the transition region. 
    Secondly, they may have dissipated below the corona. 
    However, this explanation seems to be less likely given their slow rate of dissipation due to their incompressibility. 
	Thirdly, either the spatial scale or periods of these mode waves could have been smaller (shorter) than the resolution of our observations, 
    in which case, broadening of emission line profiles occurs. 
    To investigate torsional Alfv\'{e}n mode and fast kink mode more thoroughly, 
    it is necessary to analyze in more detail the time variability and spatial distribution of emission line width. 
	However, this is beyond the scope of this paper.
	Such analysis will become possible with much higher spatial resolution. 
	Nonetheless, we feel that our method to identify MHD wave modes provides a useful diagnostic tool 
	for the statistical study of waves in the solar corona. 
    
	Hinode is a Japanese mission developed and launched by ISAS/JAXA, collaborating with NAOJ as a domestic partner, 
	NASA and STFC (UK) as international partners. 
	Scientific operation of the Hinode mission is conducted by the Hinode science team organized at ISAS/JAXA. 
	This team mainly consists of scientists from institutes in the partner countries. 
	Support for the post-launch operation is provided by JAXA and NAOJ (Japan), STFC (U.K.), NASA, ESA, and NSC (Norway).
	We greatly appreciate the proofreading/editing assistance from the GCOE program. 


\clearpage
\begin{table}
	\centering
		\caption{The correspondence between MHD mode waves and oscillations of observable quantities.} 
		\label{modes}
		\begin{tabular}{lccc} 
			\toprule
			Mode 									& Density 							& Doppler velocity		 & Line width \\ 
			\midrule
			Torsional Alfv\'{e}n mode & $0$ 		& $\neq 0$ 	& $\neq 0$ \\ 
			Fast kink mode 					& $0$ 		& $\neq 0$ 	& $0$ \\ 
			Fast sausage mode 			& $\neq 0$ 	& $0$ 			& $\neq 0$ \\ 
			Slow mode 							& $\neq 0$ 	& $\neq 0$ 	& $0$ \\ 
			\bottomrule      
		\end{tabular}
\end{table}
\clearpage
\clearpage
\end{document}